\documentclass[conference]{IEEEtran}
\IEEEoverridecommandlockouts

\usepackage{cite}
\usepackage{amsmath,amssymb,amsfonts}
\usepackage{algorithmic}
\usepackage{graphicx}
\usepackage{booktabs}
\usepackage{enumitem}
\usepackage{multirow}
\usepackage{url}
\usepackage{textcomp}
\usepackage{xcolor}
\def\BibTeX{{\rm B\kern-.05em{\sc i\kern-.025em b}\kern-.08em
    T\kern-.1667em\lower.7ex\hbox{E}\kern-.125emX}}
\begin{document}

\title{Enhancing and Exploring Mild Cognitive Impairment Detection with W2V-BERT-2.0}

\author{Yueguan Wang$^{\star \dagger}$ \qquad Tatsunari Matsushima$^{\star }$ \qquad Soichiro Matsushima$^{\star }$ \qquad Toshimitsu Sakai$^{\star \ddagger}$\\

$^{\star}$ IGSA Inc.  \quad $^{\dagger}$ The University of Tokyo  \quad $^{\ddagger}$ Toyohashi University of Technology
}






\maketitle

\begin{abstract}
This study explores a multi-lingual audio self-supervised learning model for detecting mild cognitive impairment (MCI) using the TAUKADIAL cross-lingual dataset. While speech transcription-based detection with BERT models is effective, limitations exist due to a lack of transcriptions and temporal information. To address these issues, the study utilizes features directly from speech utterances with W2V-BERT-2.0. We propose a visualization method to detect essential layers of the model for MCI classification and design a specific inference logic considering the characteristics of MCI. The experiment shows competitive results, and the proposed inference logic significantly contributes to the improvements from the baseline. We also conduct detailed analysis which reveals the challenges related to speaker bias in the features and the sensitivity of MCI classification accuracy to the data split, providing valuable insights for future research.

\end{abstract}

\begin{IEEEkeywords}
cross-lingual MCI classification, W2V-BERT 2.0, Feature analysis, Speaker bias
\end{IEEEkeywords}

\section{Introduction}

Mild cognitive impairment (MCI) is characterized by an initial decline in cognitive functions, predominantly memory and reasoning abilities. Individuals with MCI are at a higher risk of developing dementia, often caused by Alzheimer's disease (AD). On the other hand, if the patient is in the MCI stage, lifestyle modification and exercise therapy can be expected to restore cognitive function \cite{shimada2019reversible}, and effective drugs for the MCI stage \cite{van2023lecanemab} have been developed in recent years. Hence, detection of MCI is becoming increasingly important to prevent the onset and progression of dementia and to extend healthy life expectancy.

Those with MCI or AD typically have language impairments, especially at semantic and pragmatic levels \cite{stevenH13}. For example, they show difficulties finding correct words and names, invent new words, lose verbal fluency, and increase pauses during speech. Therefore, speech-based MCI and AD detection have attracted researchers' attention as a potential cost-effective screening method. Recently, the utilization of either manual or automatic speech recognition (ASR) based transcriptions and BERT pre-trained model \cite{devlin2019bert} has become a popular approach because of their competitive performance. \cite{balagopalan2020bert} fine-tune the BERT model with the classification head layer based on the manually scripted transcription, and it achieves 83.3\% accuracy for the AD detection task. \cite{syed2020automated} obtain several BERT-based models' embeddings from speech transcriptions and applied different ways of pooling. It achieves 89.81\% the highest accuracy for the AD detection task.\cite{valsaraj2021alzheimers} use auto-generated transcriptions and fine-tune a pre-trained BERT model to obtain its embeddings. The embeddings are combined with other linguistic features, which are then used for the classification. The results show the advantage of BERT features-based model over existing acoustic features-based models. Despite its potential to be be an effective approach, we observe the following limitations.

\begin{itemize}
\item The speech transcriptions are not always available. Although it is possible to obtain them using ASR, the errors from ASR will be cascaded, potentially degrading the classification performance.
\item Text does not have exact time resolution compared to speech, such as articulation speed and pauses during the speech, which are also important cues for AD/MCI detection \cite{G_mez_Zaragoz__2023}.
\end{itemize}

Motivated by these limitations, this study focuses on the investigation of the feature embeddings directly obtained from speech utterances. Since audio self-supervised learning (SSL) models have been successful in a wide range of tasks, from ASR to speech synthesis, we investigate the features, particularly from an audio SSL model. Although several studies \cite{zhang2024softweighted}, \cite{braun2023classifying}, \cite{Braun_2022}, \cite{pereztoro22_interspeech} have also explored this research line, attempts at MCI detection in a language-independent manner are limited. Therefore, the main contributions of our study are as follows.

\begin{itemize}

\item We explore using a large-scale multi-lingual audio SSL model for cross-lingual MCI classification and provide a simple but effective method to search the essential layers of audio SSL models for MCI detection. Experiments show that our selected layer is competitive. This accelerates future research by being able to get suitable features with minimal preliminary experiments. 

\item We design specific inference logic for MCI prediction, which turns out to be definitive in achieving competitive performance. 

\item We conduct detailed analysis with SSL features, with the addition of  annotated features, and show the existing challenges and promising directions to solve this challenging task. 
\end{itemize}

\section{MCI Classification Methods}
\subsection{Feature Extraction}
We consider W2V-BERT 2.0 \cite{communication2023seamlessm4t}, a large-scale multi-lingual audio SSL model for the feature extractor. W2V-BERT 2.0 was recently proposed in the SeamlessM4T model \cite{communication2023seamlessm4t} and used as a speech encoder of a translation model. 
It follows W2V-BERT \cite{chung2021w2vbert} that consists of 24 conformer layers \cite{gulati2020conformer} and combines contrastive and masked prediction learning, and it is pre-trained on 4.5M hours of unlabeled audio data that covers 143 languages.

It is known that each transformer layer in audio SSL models learns different types of features such as acoustic and semantic properties \cite{shah2021audio}, \cite{pasad2022layerwise}, \cite{Cho_2023}, \cite{lin2022utility}. Since language impairments caused by MCI mainly appear at the semantic level, it is assumed that using the layer that encodes those relevant features can improve classification performance. Generally, training separate classification models with each conformer layer's features can identify the most suitable layer given the certain task. However, it is rather time-consuming due to the large search space of large-scale SSL models. In order to tackle this challenge, we propose the following methods to get the suitable extracted features. 

\begin{figure}
    \centering
    \vspace{-5mm}
    \includegraphics[scale=0.45]{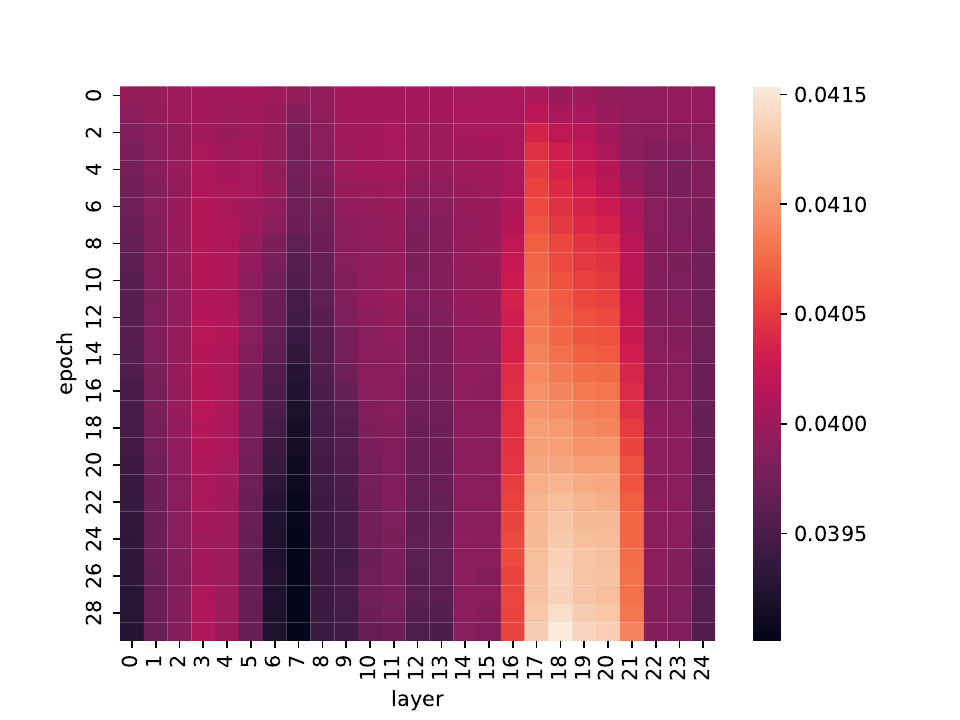}
    \vspace{-5mm}
    \caption{The visualization of how weights of layers change during the training. The weights around the 18th layer become higher gradually as training proceeds. }
    \vspace{-5mm}
    \label{uniform}
\end{figure}

\begin{description}[style=unboxed,leftmargin=0cm]
    \item[Quick Essential Layer Detection] 
 In this work, we propose a simple method to quickly figure out the essential layers given a task.  Inspired by \cite{weighted}, we leverage learnable parameters $\textbf{p}$, which assign weights to each layer's outputs, to help search the essential layers. 

First, we set the parameters with a uniform distribution and visualize how the weights change during the learning process. The results in Fig. \ref{uniform} show how the weights converge during the training. In this way, it is possible to figure out the essential layers with one single preliminary experiment.  

\item[Weighted Sum of SSL Features]On the other hand, the results of Fig. \ref{uniform} suggest that the weights can not be effectively tuned to the optimal distribution. In addition, instead of using features of single layer, we assume that combining some information from other layers could enhance the classification performance. Therefore, we give the weights a priori distribution, i.e. set a major layer with the premise that the most effective layer is already known. Concretely, the weight of $i$-th layer $p_i$ is set with:
$$
p_i =  \begin{cases}  k &   i=c \\  0.0 &  \text{other}  \end{cases}
$$
where $c$-th layer is the major layer and $k$ is the major weight ($k>0$), we set $c=18$ according to the results in Fig. \ref{uniform}. 
The final output feature $\hat{\textbf{h}}$ is then obtained by
$$
\hat{\textbf{h}} = \sigma(\textbf{p}) \cdot \textbf{h}
$$
where $\textbf{h}$ is the stack of SSL features of layers, and $\sigma(\cdot)$ is normalization function, i.e. softmax function.  
    
\end{description}

\subsection{Classifier: BiLSTM with weighted sum of SSL features}

A common assumption among past studies is that feature selection matters more than the classification model complexity or the modeling algorithm on classification performance. This is particularly reflected by the studies \cite{syed2020automated}, \cite{valsaraj2021alzheimers} that use non-neural network-based simple machine learning algorithms to achieve competitive classification accuracy. Hence, we build a simple Bidirectional LSTM (BiLSTM) network containing 5 layers with a projection layer and linear classification head on top, for the classification model. It takes $\hat{\textbf{h}}$ as input and outputs classification logits.



\section{Experiments}
\subsection{Dataset and Baseline}


We use the TAUKADIAL dataset \cite{taukadial_challenge} for experiments. This dataset contains spontaneous speech recordings of picture descriptions in English or Chinese from healthy and MCI individuals. The dataset is balanced by gender and age to reduce bias. In training set, there are 3 different audio files for each patient labelled MCI or NC (Normal Cognition) . We allocate 20\% of the training set for validation. For test, we use officially released test set, which consists of speech recordings by different patients in either of these two languages. 

Previous work related to this task \cite{luz2024connected} combines SSL features of wav2vec model \cite{schneider2019wav2vec} and eGeMAPs feature \cite{7160715}, which includes a series of acoustic features to achieve some progress in this challenging task. In our work, we use w2v (Wav2vec) + eGeMAPs as the baseline to compare with our proposed method. 


\subsection{Training and inference setup}

\subsubsection{Data preprocessing}
For the TAUKADIAL dataset, we segment the training samples every 30 seconds for actual training data to avoid challeging computation complexity of some utterances span which lasts several minutes. To remove the speaker bias, with which models tend to rely on speakers’ features for classification, we do train-validation split with patient ID to make sure patients in the two sets are completely different. Moreover, we apply information perturbation following \cite{nansy} to improve the robustness of the model against different types of speakers.
Concretely, we apply three functions: format shifting, pitch randomization, and random frequency shaped with a parametric equalizer. We follow the hyperparameters described in Appendix A of \cite{nansy}.

For the test set, we segment similarly to avoid the length discrepancy between training samples and test samples. We propose two logics on how to predict the final label with the results of split segments and discuss them on \ref{infer}. 

\subsubsection{Model training details}
For BiLSTM classifier, the hidden feature dimension and projection dimension is set to 256 and 64, with dropout set to 0.1 among all layers. During training, the parameters of the feature extraction SSL models are frozen. The model is trained with cross-entropy loss.  For major weigh $k$, we search the value from [0.0,1.0,5.0,8.0], and choose 5 based on the performance on the validation set.

All the experiments are conducted on a single RTX 4090Ti and RTX A6000 GPU. The batch size is set to 4 and the learning rate is set from [3e-5, 5e-5, 1e-4] based on the accuracy of the validation set. We use the AdamW \cite{loshchilov2018decoupled} algorithm for optimization with beta [0.9, 0.98]. To get a fair comparison among all experiments, the initial weights are fixed for all experiments. 

\begin{table*}[h]
  \caption{Results of validation set and test set under speaker (spkr) split and general data split}
  \label{split_bias}
  \centering
  \begin{tabular}{ c c | c c c c | c c c c}
    \toprule
    \multicolumn{2}{c|}{\textbf{Setting}} & \multicolumn{4}{c|}{\textbf{Validation}} & \multicolumn{4}{c}{\textbf{Test}} \\ \midrule 
     \textbf{Model}& \textbf{Split} & \textbf{ACC} & \textbf{Precision} & \textbf{Recall} & \textbf{F1} & \textbf{ACC} & \textbf{Precision} & \textbf{Recall} & \textbf{F1}\\ \midrule
    W2V + eGeMAPs & - & - & - & - & - & $0.5920$ & $0.6167$ & $0.5873$ & $0.6016$ \\
    \midrule
    \multirow{2}{*}{W2V-BERT 2.0} & spkr & $0.7194$ & $0.6718$ & $0.6837$ & $0.6777$ & $\textbf{0.6250}$ & $0.6125$ &$\textbf{0.7778}$ &$\textbf{0.6853}$ \\
    & data &  $0.9133$ & $0.8844$ & $0.9728$ & $0.9172$ & $0.6000$ & $\textbf{0.6314}$ &$0.5716$ &$0.5999$ \\ \hline
    
    \bottomrule
  \end{tabular}
\end{table*}

\begin{table}[th]
  \vspace{-5mm}
  \caption{The cross validation results of W2V-BERT 2.0 model} 
  
  \label{cv_w2vbert}
  \centering
  \begin{tabular}{ c c c c c}
    \toprule
    \textbf{Fold}  & \textbf{ACC} & \textbf{Precision} & \textbf{Recall} & \textbf{F1} \\
    \midrule
    1  & $0.7194$ & $0.6718$ & $0.6837$ & $0.6777$\\
    2 & $0.6500$ & $0.7000$ &$0.8167$ &$0.7005$ \\
    3 & $0.6200$ & $0.5333$ &$0.6400$ &$0.5439$ \\
    4 & $0.4573$ & $0.3902$ &$0.3801$ &$0.3575$ \\
    5 & $0.7744$ & $0.8293$ &$0.8720$ &$0.8276$ \\
    \bottomrule
  \end{tabular}
\end{table}

\subsubsection{Model inference logics}\label{infer}
We use the two inference logics during test time:
\begin{itemize}
    \setlength{\itemsep}{0.5pt}
    \item Ensemble Logic: The final predicted label depends on the cumulative probabilities of each candidate label. 
    \item OR Logic: The symptoms of MCI may only appear for the limited time frames in a patient's utterance, meaning most of the time frames do not have important cues for MCI detection. This could encourage mistakes in the detection even if the model is trained well. Therefore, we propose OR inference logic, which predicts MCI when at least one of the segments is predicted MCI. 
\end{itemize}


\subsubsection{Evaluation metrics}
We use four widely-used metrics: accuracy (ACC), precision, recall, and F1 score for the evaluation of our experiments.

\section{Results and Discussion}

In this section, we show the results of our proposed approaches, and discuss the conclusions drawn from experiment results. 

\subsection{Main Results}
Table \ref{split_bias} shows the results of the baseline and our approach. Our model significantly improves recall while maintaining precision, leading to improvements among most evaluation metrics compared to the baseline. The results indicate that our classification model using features of W2V-BERT 2.0 can provide competitive results in this challenging task. 

\begin{table}[th]
  \caption{The results under Ensemble and OR inference logic} 
  \label{logic}
  \centering
  \begin{tabular}{ r c c c c}
    \toprule
    \textbf{Logic}  & \textbf{ACC} & \textbf{Precision} & \textbf{Recall} & \textbf{F1} \\
    \midrule
    Ensemble  & $0.5500$ & $0.5714$ & $0.5714$ & $0.5714$\\
    OR & $\textbf{0.6250}$ & $\textbf{0.6125}$ &$\textbf{0.7778}$ &$\textbf{0.6853}$ \\
    \bottomrule
  \end{tabular}
\end{table}

\subsection{Discussion}
\begin{figure}[h]
    \hspace{-5.2mm}
    \setlength{\belowcaptionskip}{-8pt}
    \includegraphics[width=9cm]{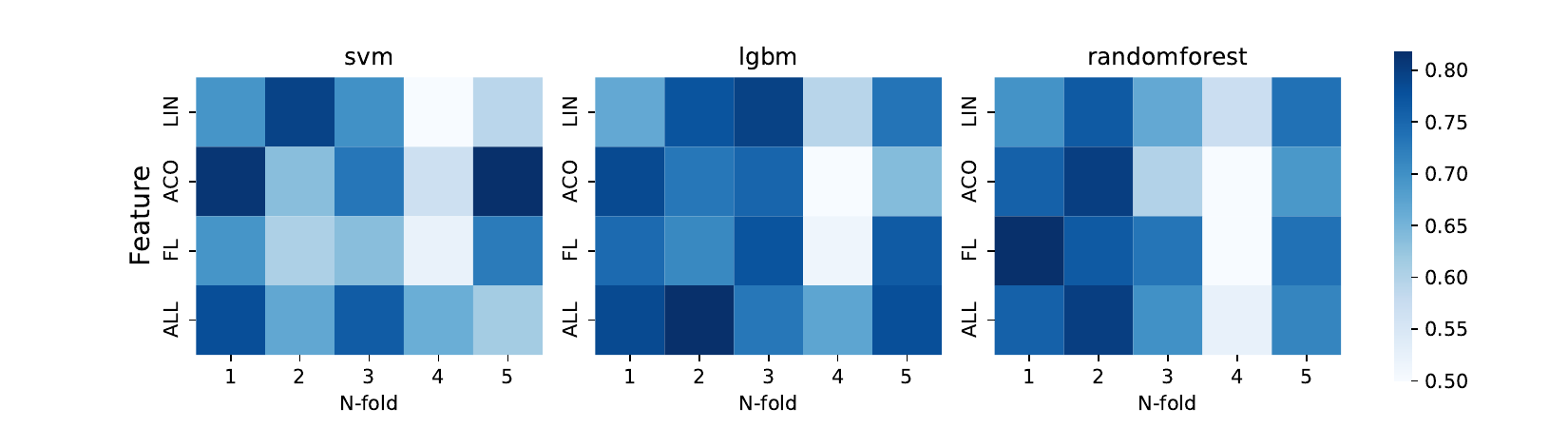}
    \caption{Accuracy results of annotated features for each fold. LIN, ACO, FL and ALL denote linguistic, acoustic, fluency and all features, respectively. }
    \label{ml}
\end{figure}

\begin{figure}
    \vspace{-4.2mm}
    \centering
    \includegraphics[width=8cm]{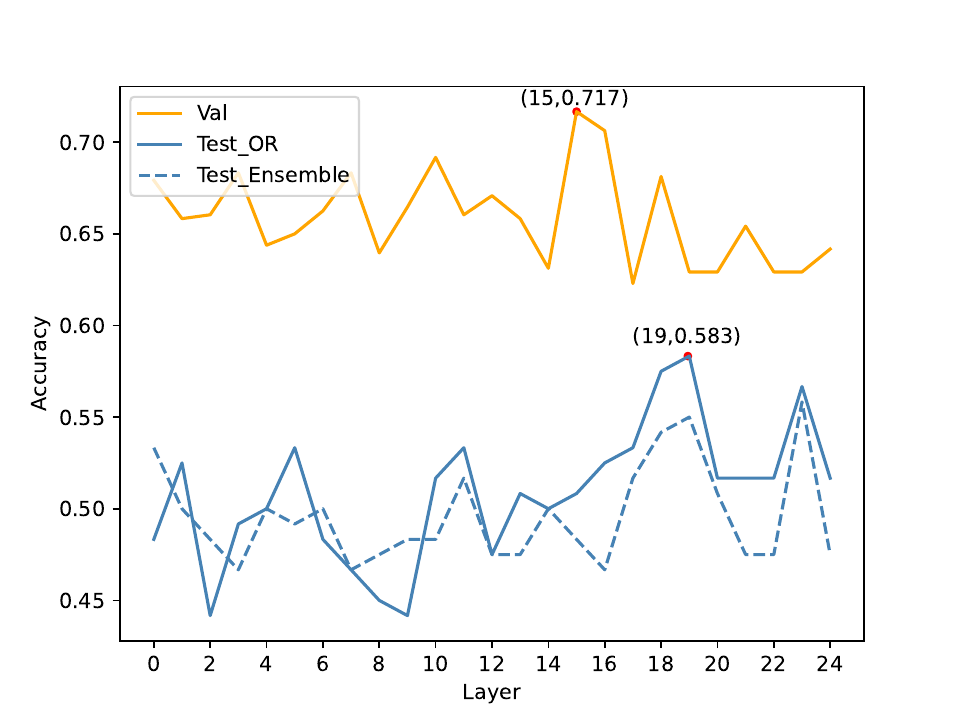}
    \caption{Accuracy of validation set and test set under different inference logics with features of single layer. The peaks are marked as \((layer_{id}, acc\)). }
    \label{layer_res}
\end{figure}

\begin{figure}[h]
    \vspace{-5.2mm}
    \centering
    \includegraphics[scale=0.5]{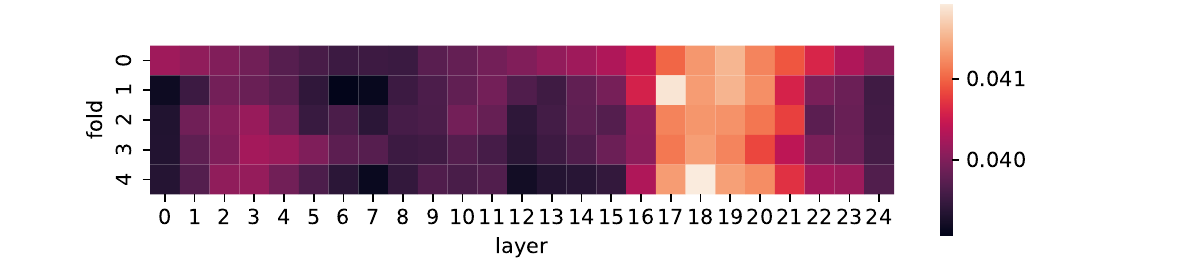}
    \vspace{-8.2mm}
    \caption{The layer weights of the final epoch in cross validation folds. }
    \vspace{-3.0mm}
    \label{unifold}
\end{figure}

\begin{description}[style=unboxed,leftmargin=0cm]
    \item[Effectiveness of Essential Layer Detection and Weight Sum of SSL features] To demonstrate the effectiveness of our major layer decision method and the weighted sum of W2V-BERT 2.0 features, we train a classifier with each layer's features of W2V-BERT 2.0. The results of the validation and test set are shown in Fig. \ref{layer_res}. The peak of validation and test accuracy both appear at the interval between layers 15 and 19, which is consistent with the visualization of Fig. \ref{uniform}. Also, our choice, the 18th layer, provides competitive accuracy with both evaluation sets. 

In addition, by referring to the results of Table \ref{split_bias}, compared to using features of a single layer, the classifier with the weighted sum of features achieves better accuracy on both sets, which indicates combining features of multiple layers is beneficial. On the other hand, we initialize the weights with only one major layer, and the optimization of initialization remains to be explored in the future.  

\item[How do different inference logic affect results?]
We show the results of the test set under the two proposed inference logics. As Table \ref{logic} shows, OR logic significantly increases the recall while even slightly increasing the precision. Concretely, it corrects 13 false negatives while only 4 true negatives are mistaken. It reveals that most segments of some MCI samples are predicted as NC and only a few are predicted as MCI, and OR logic corrects these samples. On the other hand, for audios with NC true label, all segments tend to be predicted NC, thus less effected under OR logic. The single layer results in Fig. \ref{layer_res} also show that OR logic superiors in all valid cases, indicating the robustness of OR logic. 

\item[Speaker Information Bias of W2V-BERT 2.0 Features]
In our main experiments, we split the data by patients to remove the speaker bias. We adopt the general data split with the same split ratio and show how the results differ depending on the split method in Table \ref{split_bias}. Although the bias issue is partly mitigated under information perturbation, with the general data split, the accuracy of the validation data significantly increases while the results of test data are maintained. This indicates that classifiers tend to leverage speaker information under general data split. However, some work using the same dataset such as \cite{cheng24c_interspeech}, has ignored this bias.  

We note that this bias can not be completely removed even under speaker split and information perturbation augmentation, according to the differences between the results on the validation and test set. And classification results are still affected by this bias. This suggests the necessity to extract more reliable and fair features in future work.

\item [Sensitivity of Data Split] We explore the data sensitivity of our proposed approach by conducting cross validation experiments.  
\subsubsection{Cross Validation with SSL features}
To figure out whether this approach is robust to data split, we conduct cross-validation experiments with 5 folds considering that we use 20\% of the whole data for validation. The first fold is the same as the split of the main experiments. Table \ref{cv_w2vbert} shows the results of the validation set under the different cross-validation results. The significant variance among those 5 folds indicates the data sensitivity of our approach. 

\subsubsection{Cross Validation with Manually Annotated features}

Although past study \cite{stevenH13} shows that MCI or AD tend to have language impairments at semantic and pragmatic levels, it is still important to investigate prominent features in spoken utterances. To figure out how W2V-BERT 2.0 features align with existing features that are easily interpretable, i.e. acoustic, fluency and linguistic features, following \cite{10031189} and \cite {balagopalan2020bert}, we also conduct cross-validation of machine learning methods with manually annotated features under the same 5 folds. Concretely, we extract speech features that are categorized into fluency, linguistic and acoustic features and conduct classification experiments with three machine learning models, LightGBM\cite{lgbm}, RandomForest \cite{randomforest} and SVM \cite{SVM}. The hyperparameters of models are optimized with optuna\cite{optuna}. Fig. \ref{ml} shows the cross validation results of 5 folds. 

All models' cross-validation results show consistent trend, with the fourth fold's performance significantly lower than other folds. The consistency shows the alignment between the contained information of SSL features and annotated features. The high variance among cross-validation folds indicates that the MCI classification is sensitive to data splits in general, highlighting the difficulty of the task and the necessity of developing a more robust feature extractor for a large and diverse patient population.

\subsubsection{Robustness of Quick Essential Layer Detection} According to Fig. \ref{unifold}, the weight distribution is stable among 5 folds, indicating essential layer detection is robust enough for data split, which is effective for searching the essential layer given a certain task.
\end{description}

\section{Conclusion}
This study explores a multi-lingual audio SSL model, specifically W2V-BERT-2.0 for cross-lingual MCI detection. Given the challenge of searching essential layers, we propose a simple but effective essential layer selection method by visualizing the weights of individual layers during the training. We also propose an inference logic specifically designed for MCI classification, based on the assumption that important cues for MCI only appear in a part of an utterance. Experiment results show that our choice of essential layer achieves competitive results, which slightly surpasses the baseline model. In addition, we show that our inference logic plays a crucial role in performance improvements. 

However, our ablation studies also show some limitations on the robustness of the proposed method. With a data split ablation study, we show that W2V-BERT 2.0 features still contain speaker bias regardless of the attempt of the data augmentation to remove the speaker characteristics. Additionally, we reveal the challenge of MCI classification with W2V-BERT 2.0 features and manually annotated features. Regardless of classification models, the significant performance difference among cross-validation folds shows the high sensitivity to data split and indicates the difficulty of the task. 

Future research will focus on the exploration of more robust and fair features of pre-trained SSL models to further improve this little-explored yet essential task. 

\bibliographystyle{IEEEtran}
\bibliography{strings,refs}

\end{document}